\documentclass[12pt]{iopart}
\usepackage{epsfig}
\usepackage{iopams}

\newcommand{\be}{\begin{equation}}
\newcommand{\ee}{\end{equation}}

\newcommand{\ba}{\begin{array}}
\newcommand{\ea}{\end{array}}
\newcommand{\bea}{\begin{eqnarray}}
\newcommand{\eea}{\end{eqnarray}}
\newcommand{\qed}{\begin{flushright} $\square$
                  \end{flushright}
}
\newcommand{\g}{\mathfrak{g}}
\newcommand{\nn}{\nonumber \\}

\newcommand{\Q}{R}

\newtheorem{theorem}{Theorem}
\newtheorem{corollary}{Corollary}
\newtheorem{proposition}{Proposition}

\begin{document}

\title{Soliton surfaces via zero-curvature representation of differential equations }
\author{ A M Grundland$^1$ $^2$ and S Post$^1$}

\address{$^1$ Centre de Recherches Math\'ematiques. Universit\'e de Montr\'eal. Montr\'eal CP 6128 (QC) H3C 3J7, Canada}
\address{$^2$ Department of Mathematics and Computer Sciences, Universit\'e du Quebec, Trois-Rivi\`eres. CP500 (QC)G9A 5H7, Canada}
\ead{grundlan@crm.umontreal.ca, post@crm.umontreal.ca}

\begin{abstract} The main aim of this paper is to introduce a new version of the Fokas-Gel'fand formula for immersion of soliton surfaces in Lie algebras. The paper contains a detailed exposition of the technique for obtaining exact forms of 2D-surfaces associated with any solution of a given nonlinear ordinary differential equation (ODE) which can be written in zero-curvature form. That is, for any generalized symmetry of the zero-curvature condition of the associated integrable model, it is possible to construct soliton surfaces whose Gauss-Mainardi-Codazzi equations are equivalent to infinitesimal deformations of the zero-curvature representation of the considered model. Conversely, it is shown (Proposition 1) that for a given immersion function of a 2D-soliton surface in a Lie algebra, it possible to derive the associated generalized vector field in evolutionary form which characterizes all symmetries of the zero-curvature condition. The theoretical considerations are illustrated via surfaces associated with the Painlev\'e equations P1, P2 and P3, including transcendental functions, the special cases of the rational and Airy solutions of P2 and the classical solutions of P3. 
  \end{abstract}

\section{Introduction}
The Fokas-Gel'fand formula for immersion associated with integrable models is a powerful tool for the construction and investigation of 2D soliton surfaces in Lie algebras \cite{FG, FGFL}. In \cite{GP2011a}, the authors described how a zero-curvature representation of integrable nonlinear partial differential equations (PDEs) can be used, via the Fokas-Gel'fand formula,  to induce surfaces immersed in Lie algebras via symmetries of the compatibility conditions for a linear spectral problem (LSP). The procedure was then applied to integrable nonlinear ordinary differential equations (ODEs) admitting a Lax representation \cite{GP2011b}. The results obtained were so promising that it seemed to be worthwhile to try to extend this method and check its effectiveness for a different form of a zero-curvature condition (ZCC). This is, in short, the aim of the present paper which provides a self-contained comprehensive study of the symmetry approach to the Fokas-Gel'fand immersion formula as applied to both PDE and ODE cases. Namely, the general theory is developed for each of the following three forms of matrix LSPs: zero-curvature representation (ZCR) of PDEs, Lax representation of ODEs and ZCR of ODEs. In addition to more general considerations, we focus on the third form of a ZCC for a differential equation, that of zero-curvature representations for ODEs involving the differentiation of the potential matrices in a Lax pair with respect to the independent variable and the spectral parameter. Such a Lax pair has  been used in the study of Painlev\'e equations. 

Each of the three types of Lax pair described above, can be thought of as a sub-case of a general ZCC and to this end, we describe infinitesimal deformations of such equations and their associated surfaces. The zero-curvature representation of the Gauss-Mainardi-Codazzi (GMC) equations  is a well known tool for the analysis of many, varied forms of integrable surfaces \cite{ BavMarv2010, Bob1994, CGS1995}.
The necessary and sufficient conditions for the existence of surfaces immersed in the Lie algebra $\g$ whose GMC equations are equivalent to an infinitesimal deformation of the ZCC are given in terms of the determining equation for generalized symmetries of the ZCC.  For this purpose, in Proposition 1,  we give a complete classification of all generalized symmetries of the ZCC via an isomorphism between such symmetries and surfaces written in the moving frame defined by conjugation by the wave function in the LSP. 

In addition to symmetries of the ZCC, it is possible to obtain infinitesimal deformations of the ZCC from the integrable model considered, see \cite{FGFL}. In \cite{GP2011a, GP2011b}, the necessary and sufficient conditions for the existence and explicit integration of the surfaces were given in terms of the symmetry criterion for generalized vector fields. Taken together, the symmetries of the ZCC and the integrable model give a quite general form for soliton surfaces, which we refer to as the Fokas-Gel'fand formula for  immersion of surfaces in Lie algebras. These surfaces include those already known, namely those constructed from a conformal symmetry of the spectral parameter (the Sym-Tafel formula for immersion \cite{Sym, Tafel}), gauge symmetry \cite{Cies1997, FGFL}, and generalized symmetries of the integrable system and its LSP \cite{FGFL}. In this paper, we give additional generalized symmetries of the ZCC and their associated soliton surfaces. 

In many cases, including those considered herein, the classical symmetry analysis is not a proper tool for obtaining these types of surfaces because the Lie point symmetries of the initial system, written in terms of the ZCC and its LSP, are too restrictive. To overcome this difficulty, we make use of the concept of a generalized symmetry as introduced by E. Noether \cite{Noether}. The progress in studying general properties of generalized symmetries has been subsequently developed by several authors (see e.g. \cite{Olver} and references therein). The most important element of this approach is the introduction of a symmetry formalism of generalized vector fields (defined on some extended jet space), their prolongation structure and their links with the Fr\'echet derivative.

An interesting consequence of the adopted approach is that it leads to a direct connection between generalized symmetries and their associated surfaces for Painlev\'e type equations (see e.g. \cite{Incebook}). We focus here on the Painlev\'e equations P1, P2 and P3 as examples, namely
\bea P1: & x_{tt}=6x^2+t,\nn
P2: & x_{tt}=2x^3+tx-\alpha,\nn
P3: & x_{tt}=\frac{(x_t)^2}{x}-\frac{x_t}{t}+\frac1t(\alpha x^2+\beta)+\gamma x^3+\frac{\delta}{x}.\nonumber \eea
The quantities $\alpha, \beta, \gamma, $ and $\delta$ in the above equations are free parameters. For specific values of these parameters, several explicit solutions of the Painlev\'e equations have been found (see e.g. for review \cite{AblClabook, Conte1999,  FIKNbook, Gromak1999, Iwaetal1991, McLOlv1983}). The successive application of B\"acklund transformations (with different values of parameters) allows one to create new solutions of a Painlev\'e equation from the old ones. In this context, the symmetry analysis of Painlev\'e equations has been systematically undertaken in \cite{Noumibook} and has proven to be a useful tool since it leads to new solutions. However, the present state of solvability of the Painlev\'e equations through the so-called Painlev\'e transcendents is still not satisfactory from the point of view of symmetry groups. The progress in studying general properties of generalized symmetries of Painlev\'e equations and their LSPs opens a possibility for applying this symmetry approach to the construction of soliton surfaces. 
 
 An original procedure for constructing surfaces associated with Painlev\'e equations via isomonodromic deformations was devised by A. Bobenko and U. Eitner in \cite{BobEitbook}. These equations, written in zero-curvature form involving the differentiation of the potential matrices in a Lax pair with respect to the independent variable and the spectral parameter, arise from the  compatibility conditions of an LSP.  The surfaces considered in this work correspond to infinitesimal deformations of such surfaces as the corresponding structural equations are infinitesimal deformations of the zero-curvature conditions of the integrable models.

This paper is organized as follows. In section 2, we discuss surfaces immersed in Lie algebras whose GMC equations are given by infinitesimal deformations of the ZCC. 
In  section 3, we discuss three different ways that the ZCC can be realized as the compatibility condition for the LSP of an integrable model and how the symmetries of this model give further surfaces. The case where the ZCC is realized by the Lax pair for an ODE is given in detail in section 4. To illustrate the theoretical considerations, section 5 contains explicit construction and geometric analysis of surfaces associated with the Painlev\'e equations P1, P2 and P3.  Section 6 contains final remarks and possible future developments.

\section{The zero curvature condition and its associated surfaces}
\subsection{The ZCC and its jet space}
Consider the following partial differential equation (PDE) which, in what follows,  shall be referred to as the zero-curvature condition (ZCC), 
\be \label{Delta} \Delta[u]=D_2 u^1-D_1 u^2 +\left[u^1, u^2\right]=0,\ee
with independent variables $\xi_{i}, \ i=1,2$ and dependent matrix variables $u^\alpha,$ $\alpha=1,2$ which take their values in a Lie algebra $\g.$ Here, we write \eref{Delta} as a function on the jet space  $M\equiv(\xi_1, \xi_2, u^\alpha, u^\alpha_J),$  where  the derivatives of $u^\alpha$ are given by 
 \be \label{coordsjetspace}\frac{\partial ^n}{\partial \xi_{j_1}\ldots \partial \xi_{j_n}} u^\alpha\equiv u^\alpha_J, \qquad J=(j_1, \ldots j_n), \quad j_i=1,2, \ |J|=n.\ee 
 We define $\mathcal{A}\equiv C^\infty (M)$ to be the set of smooth functions on the  jet space $M$ and  use the abbreviated notation $f(\xi_1, \xi_2, u^\alpha, u^\alpha_J)\equiv f[u]\in\mathcal{A}.$ Thus, the PDE \eref{Delta} becomes a vanishing element of $\mathcal{A}.$
 The derivative of a function on the jet space $M$, in the direction $\xi_i,$ is given by the total derivative $D_i$ defined as 
 \be D_i=\frac{\partial}{\partial \xi_i}+u^{\alpha}_{J,i}\frac{\partial}{\partial u^\alpha_J}, \qquad \alpha,i=1,2.\ee
The  ZCC \eref{Delta} can be realized as the compatibility conditions of an associated linear spectral problem (LSP) 
 \be\label{LSP} D_\alpha \Psi=u^\alpha \Psi, \qquad \alpha=1,2,\ee
 where $\Psi=\Psi([u])$ is a function from the jet space $M$ to a Lie group, G, whose algebra is $\g$.
  Note that $\Psi$ satisfying the LSP \eref{LSP} is an immersion function for a surface immersed in the Lie group $G$ where the compatibility of the tangent vectors $D_\alpha \Psi$ is equivalent to the ZCC \eref{Delta}. Furthermore, for any (generalized) symmetry of \eref{Delta} there exists a $\g$-valued immersion function, $F[u],$  defined in terms of its tangent vectors
  \be \label{DF} D_1 F=\Psi^{-1} Q^1\Psi, \qquad D_2 F=\Psi^{-1} Q^2\Psi,\ee
  where $Q^1$ and $Q^2$ satisfy an infinitesimal deformation of the ZCC \eref{Delta}, namely 
 \be\label{deteq} D_1 Q^2-D_2Q^1+[Q^1,u^2]+[u^1, Q^2]=0.\ee
 In what follows, we suppose that $F$ is sufficiently smooth. 
 The existence of such surfaces as well as the determining equation \eref{deteq} was first identified for point symmetries by Fokas and Gel'fand in  \cite{FG}.
 
 Following the notation as in \cite{Olver}, a generalized symmetry of \eref{Delta} is given by a vector field, in evolutionary form, 
 \be \vec{v}_Q=Q^{\alpha,j}[u] \frac{\partial}{\partial u^{\alpha,j}},\ee
where $u^{\alpha, j}$ are the components of  $u^\alpha$ in a basis for $\g,$ i.e.  $u^{\alpha}=u^{\alpha, j}e_j$ with $j=1...n.$ Note that $Q^\alpha\equiv Q^{\alpha, j}e_j$ is an element of the Lie algebra $\g$. The prolongation of $\vec{v}_Q$ is defined to be 
 \be pr\vec{v}_Q=\vec{v}_Q+D_J(Q^{\alpha,j}[u]) \frac{\partial}{\partial u^{\alpha,j}_J}.\ee
 The vector field $\vec{v}_Q$ is a generalized symmetry of a nondegenerate PDE \eref{Delta} if and only if \cite{Olver}
 \be pr\vec{v}_Q(\Delta[u])=0, \qquad \mbox{ whenever } \Delta[u]=0,\ee
 holds. In fact the prolongation of $\vec{v}_Q$ acting on $\Delta[u]=0$ is exactly  the determining equations \eref{deteq}. We have the following theorem. 
 
 \begin{theorem}[Existence of the $\g$-valued immersion function] Suppose that there exists a smooth, $G$-valued function $\Psi$ which satisfies the LSP \eref{LSP}. Suppose also that   the generalized vector field, $\vec{v}_Q,$ is a symmetry of the ZCC \eref{Delta}. Then there exists a smooth $\g$-valued function  $F[u]$ with tangent vectors given by \eref{DF}. Further, these tangent vectors are explicitly integrated  as 
 \be \label{F} F=\Psi^{-1} pr\vec{v}_Q\Psi\in \g\ee
 if and only if the vector field $\vec{v}_Q$ is a symmetry of the LSP \eref{LSP}  in the sense that 
 \be\label{deteqLSP} pr\vec{v}_\Q\left(D_\alpha\Psi-u^\alpha \Psi\right)=0, \qquad \mbox{ whenever } D_\alpha\Psi-u^\alpha\Psi=0.\ee
 \end{theorem}
 This theorem has been proved in \cite{GP2011a}. It need only be observed that  $Q^\alpha=pr\vec{v}_Q(u^\alpha)$ and that the compatibility conditions for \eref{DF} are exactly the determining equations \eref{deteq}. The latter are in turn equivalent to $pr\vec{v}_Q(\Delta[u])=0$ since the total derivatives $D_\alpha$ commute with the prolongation of vector fields in evolutionary form, $pr\vec{v}_Q$, (see Lemma 5.12 page 306 in \cite{Olver}). To prove that $F$ given by \eref{F} has the appropriate tangent vectors, one need only compute the tangent vectors from \eref{F} to show that they coincide with \eref{DF} if and only if \eref{deteqLSP} holds.
 
Note that whenever the characteristics $Q^1$ and $Q^2$ are linearly independent and the tangent vectors $D_1F$ and $D_2F$ are compatible, the  differential 1-form $dF$ is closed  and its integral 
\be\label{dF} F=\int_\gamma \Psi^{-1}Q^1\Psi d\xi_1+\Psi^{-1} Q^2\Psi d\xi_2\in \g,\ee
 depends only on the end points of the trajectory $\gamma$ in the plane $\mathbb{R}$. The integral \eref{dF} defines a mapping $F:\mathbb{R}\backepsilon (\xi_1, \xi_2)\rightarrow \g$ which coincides with \eref{F} whenever \eref{deteqLSP} hold. The representation of a surface in terms of the 1-form \eref{dF} coincides with the generalized Weierstrass formula for immersion as introduced in \cite{Kono1996}. 
  
\subsection{Symmetries of the ZCC and associated surfaces}\label{symalpha}
In this section, we present the symmetries of the ZCC and their associated surfaces. Note that Theorem 1 says that from a given symmetry of the ZCC it is possible to construct a surface immersed in the Lie algebra $\g$. Next, we shall prove the converse. That is given a surface immersed in the Lie algebra $\g$,  it is possible to define symmetries of the ZCC and, as a result, to characterize all generalized symmetries of the ZCC in terms of a gauge function. Finally, one gets the following. 

\begin{proposition}[Extracting symmetries from a surface] Suppose that $\Psi[u]$ is a smooth, $G$-valued solution of the LSP \eref{LSP} for any solution $u^\alpha$ of the ZCC $\Delta[u]=0$. Suppose further that  $F[u]$ is a smooth function on jet space $M$ taking values in the Lie algebra $\g$, then the generalized vector field in evolutionary form 
\be  \vec{v}_Q=Q^{\alpha,j}[u] \frac{\partial}{\partial u^{\alpha,j}},\qquad \label{QsF} Q^{\alpha}=\Psi\left( D_\alpha F\right)\Psi^{-1},\quad \alpha=1,2 \ j=1,\ldots, n\ee
is a symmetry of the ZCC, $\Delta[u]=0.$ 
\end{proposition}
{\bf Proof}: Since the components of $F[u]$ are smooth functions, the cross partials commute, i.e. 
$D_1D_2F=D_2D_1F$. Thus, $Q^1$  and $Q^2$ defined in terms of $F[u]$ as in \eref{QsF} satisfy 
\bea D_2 Q^1&=&u^2\Psi( D_1F)\Psi^{-1} +\Psi( D_2D_1F)\Psi^{-1} -\Psi( D_1F)\Psi^{-1} u^2\nn
 					  &=& u^2Q^1-Q^1u^2+\Psi (D_2D_1F)\Psi^{-1},\eea
 and similarly, 
 \be D_1Q^2=u^1Q^2-Q^2u^1+  \Psi( D_2D_1F)\Psi^{-1}.\ee
 Hence, $Q^1$ and $Q^2$ defined as in \eref{QsF} satisfy 
 \be D_1 Q^2-D_2Q^1+[Q^1,u^2]+[u^1, Q^2]=pr\vec{v}_Q(\Delta[u])=0\ee
 for any solutions of the ZCC $\Delta[u]=0.$ \qed

\begin{corollary} Any generalized symmetry of the ZCC $\Delta[u]=0$ can be written in terms of a gauge function $S[u] \in \g$  as
 \be\label{QsS} Q^1=D_1 S+[S, u^1], \qquad Q^{2}=D_2 S+[S, u^2].\ee
 \end{corollary}
 {Proof:} Since $\Psi$ is invertible, it is possible to express any $\g$-valued function $F[u]$ in terms of a gauge matrix $S[u]$
 \be F[u]=\Psi^{-1} S[u] \Psi, \qquad D_\alpha F[u]=\Psi^{-1}\left(D_\alpha S[u] +[S[u], u^\alpha]\right)\Psi, \ee
 so that the characteristic $Q^\alpha$ of the vector field $\vec{v}_Q$, defined as in \eref{QsF}, satisfies \eref{QsS}.  \qed
 
From this corollary, one can construct an infinite set of generalized symmetries. As examples we give the following symmetries, beginning with the most general
\be \fl  \ba{lr} 
\vec{v}_{Q_0}=\left(D_\alpha(S)+\left[S,u^\alpha\right]\right)^j\frac{\partial}{\partial u^{\alpha j}}&\!\!\!\!\!\!\!\! \!\!\!\!\!\!\!\!\!\!\!\!\!\!\!\!\!\!\!\mbox{ gauge transformation of the LSP }\\
 \vec{v}_{Q_i}=u^{\alpha j}_i\frac{\partial}{\partial u^{\alpha j}}&\!\!\!\!\!\!\!\!\!\!\!\!\!\!\mbox{  translations in } \xi_i, \ i=1,2 \\
\vec{v}_{Q_3}=D_1\left(\xi_1 u^1\right)\frac{\partial}{\partial u^1}+\xi_1D_1\left( u^2\right)\frac{\partial}{\partial u^2}&\!\!\!\!\!\!\!\!\mbox{ dilation in } \xi_1, u^1\\
\vec{v}_{Q_4}=\xi_2 D_2\left( u^1\right)\frac{\partial}{\partial u^1}+D_2\left(\xi_2 u^2\right)\frac{\partial}{\partial u^2}&\!\!\!\! \! \!\!\!\mbox{ dilation in } \xi_2, u^2\\
\vec{v}_{Q_5}=\left(D_1^2u^1+[D_1u^1,u^1]\right)\frac{\partial}{\partial u^1}+ \left(D_1^2u^2+[D_1u^2,u^1]\right)\frac{\partial}{\partial u^2}&\!\!\! \!\!\!\!\mbox{generalized symmetry}.\ea \ee

In each of these cases, it is possible to give the explicit form of the function $F$ with tangent vectors as in \eref{DF} as was demonstrated in \cite{FG}. In what follows, we assume that the tangent vectors listed below are linearly independent. Then, $F^{i}$ for $i=1,\ldots 5$ are the immersion functions of 2D-surfaces in the Lie algebra $\g$.  
The surface associated with the gauge symmetry (studied in \cite{Cies1997, FGFL}) is, in terms of $S[u]\in g$,
\bea F^{S}=\Psi^{-1}S\Psi, \\
D_\alpha F^{S}=\Psi^{-1}\left( D_\alpha S+[S, u^\alpha]\right)\Psi\nonumber.\eea
As noted above, $F^{S}$ is in fact an arbitrary surface immersed in the Lie algebra $\g$ and the gauge function $S[u]$ can be thought of as the representation of the surface in the moving frame defined by conjugation with respect to $\Psi\in G.$

Below, we give several examples of symmetries of the ZCC \eref{Delta} and the corresponding surfaces, or equivalently the gauges, to which they are associated. The surfaces associated with the symmetries $\vec{v}_{Q_1}$ and $\vec{v}_{Q_3}$ are special cases of a more general surface given, in terms of $r(\xi_1),$ an arbitrary function of $\xi_1,$ by 
\bea F^{r}=r(\xi_1) \Psi^{-1} D_1 \Psi,\\
 D_1 F^{r}=\Psi^{-1}D_1\left(r(\xi_1)u^1\right) \Psi,\qquad 
 D_2 F^{r}=r(\xi_1)D_1(u^2)\Psi.\eea
  The gauge for this surface  is 
\[ S^{r}[u]=\Psi F^{r}\Psi^{-1}=r(\xi_1)u^1,\]
and  the characteristics $Q^\alpha$ for the vector field $\vec{v}_Q$ associated with this gauge are
\bea\fl  Q^{1}=D_1S^{r}+[S^r,u^1]=D_1\left(r(\xi_1)u^1\right),\qquad
         Q^{2}=D_2 S^{r}+[S^r,u^2]=r(\xi_1)D_1(u^2). \eea
 In the case that $r(\xi_1)=1$, the vector field $\vec{v}_{Q}$ reduces to $\vec{v}_{Q_1}$ and the surface is given by 
 \[ F^{1}=\Psi^{-1} D_1 \Psi.\]
 If instead the function reduces to $r(\xi_1)=\xi_1$, the vector field becomes $\vec{v}_{Q_3}$ and the surface is 
 \[ F^{3}=\xi_1 \Psi^{-1} D_1 \Psi.\]
 
Similarly, the surfaces associated with the symmetries $\vec{v}_{Q_2}$ and $\vec{v}_{Q_4}$are special cases of the more general surface,  
\bea
 F^{s}=s(\xi_2) \Psi^{-1} D_2 \Psi,\\
D_1 F^{s}=s(\xi_2)\Psi^{-1}D_2u^1 \Psi,\qquad \nonumber
D_2 F^{s}=\Psi^{-1}D_2\left(s(\xi_2)u^2\right) \Psi
,\eea
where $s(\xi_2)$ is an arbitrary function of $\xi_2.$ The gauge associated with this surface is 
\[ S^{s}[u]=\Psi F^{s}\Psi^{-1}=s(\xi_2)u^2,\]
and  the characteristics $Q^\alpha$ for the vector field $\vec{v}_Q$ which give this particular gauge are 
\bea Q^{1}=D_1S^{s}+[S^s,u^1]=s(\xi_2)D_2\left(u^1\right),\nn
         Q^{2}=D_2 S^{s}+[S^s,u^2]=D_2\left(s(\xi_2)u^2\right)\label{vecST}. \eea
   The construction of the surfaces $F^2$ and $F^4$ is similar to that above for the surfaces $F^1$ and $F^3$ with $\xi_1$ replaced with $\xi_2.$

The point symmetries associated with the vector fields $\vec{v}_{Q_1}, \ldots, \vec{v}_{Q_4}$ (translations and scalings) and the  surfaces $F^{1}, \ldots , F^{4}$ were identified in \cite{FG}. 

 The final symmetry to be considered as an example is a generalized symmetry, 
\bea F^{5}=\Psi^{-1}D_1u^1\Psi,\\
\label{D1F6}\fl D_1 F^{5}=\Psi^{-1}\left(D_1^2u^1+[D_1u^1,u^1]\right)\Psi,\qquad 
  D_2 F^{5}= \Psi^{-1}\left(D_1^2u^2+[D_1u^2,u^1]\right)\Psi.\eea
It is immediate to see that the tangent vectors to $F^{5}$ satisfy \eref{D1F6} . Perhaps less obvious is that the vector field, 
\be \label{v6} \vec{v}_{Q_5}=\left(D_1^2u^1+[D_1u^1,u^1]\right)\frac{\partial}{\partial u^1}+\left(D_1^2u^2+[D_1u^2,u^1]\right)\frac{\partial }{\partial u^2},\ee
is in fact a generalized symmetry of $\Delta[u]=0.$
Computing the action of the  prolongation of  the vector field $\vec{v}_{Q_5}$  on $\Delta[u]=0$, in the coordinates of the extended jet space \eref{coordsjetspace}, gives
\bea \fl pr\vec{v}_{Q_5}(\Delta[u])&=&u^1_{112}+[u^1_{12},u^1]+[u^1_1, u^1_2]-u^2_{111}-[u^2_{11},u^1]-[u^2_1,u^1_1]\nn
\fl && +[u^1_{11},u^2]+\left[[u^1_1,u^1],u^2\right]+[u^1,u^2_{11}]+\left[u^1, [u^2_1,u^1]\right]\nn
\fl &=&D_1^2(\Delta[u])+\left[D_1(\Delta[u]),u^1\right]-\left[\Delta[u],u^1_1\right]\nn
\fl &=& 0, \qquad \qquad \qquad \qquad  \mbox{ whenever } \Delta[u]=0\nonumber.\eea
Thus, the vector field $ \vec{v}_{Q_5}$ \eref{v6} is a generalized symmetry of $\Delta[u]=0.$ 

In this section, we have shown how generalized symmetries of the ZCC \eref{Delta} can be used to construct surfaces whose GMC equations are given by infinitesimal deformations of the ZCC \eref{Delta}. In the process, we were able to characterize all possible symmetries of the ZCC \eref{Delta} in terms of a gauge transformation of the associated LSP.  In the next section, we shall consider cases where the ZCC \eref{Delta} is equivalent to some integrable model and the surfaces associated to such a model. In this case, the form of surfaces obtained is enlarged since the considered symmetries are not only those of the ZCC \eref{Delta} but also of the integrable model itself. In general, these two classes of symmetries are distinct.

 \section{Integrable equations and their Lax pairs}\label{integrable}
Consider a Lax pair representation of an integrable model whose determining equation is equivalent to the ZCC \eref{Delta}. That is, suppose that we are able to parameterize the matrices  \[\tau(u^\alpha)= U^\alpha[\theta]\in \g\] in terms of some set of dependent variables $\theta^j$ which depend on (possibly a subset of)  the independent variables  $\xi^j$. The matrices  $U^\alpha[\theta]$ are also allowed to depend on the independent variables themselves, $\xi^i$ and possibly some space of constants. Define $N$ to be the jet space associated with $\theta^j$ and its derivatives and $\mathcal{B}$ to be the space of smooth functions on this jet space, which depend also on a finite number, say $k,$ of constants that take values in some field $\mathbb{K}$, usually either $\mathbb{C}$ or $\mathbb{R}.$ 

 This parameterization induces a mapping between function spaces $\mathcal{A} $ and $\mathcal{B}$
 \be \begin{array}{ccc} 
  & \tau &\\
 u^\alpha\in \g &\rightarrow& U^\alpha([\theta])\in \g \\
 \downarrow & & \downarrow\\
 & \tau &\\
 \mathcal{A}=C^\infty(M) & \rightarrow &\mathcal{B}=C^\infty(N\times \mathbb{K}^k), \end{array} \ee
 where the function $\tau$ is defined by the mapping $u^\alpha_J\rightarrow D_J U^\alpha[\theta]$ so that 
 \be \tau\left( f(\xi_1,\xi_2,u^\alpha, u^\alpha_J)\right)=f\left(\xi_1, \xi_2, U^\alpha, D_JU^\alpha[\theta] \right)\in \mathcal{B}.\ee

Under this mapping, the ZCC \eref{Delta} is converted into a system of differential equations for the dependent variables $\theta^j$ denoted 
\bea \label{Delta'} \Omega[\theta]&=&\tau(\Delta[u])\\
  &=&D_1U^2[\theta]-D_2U^1[\theta]+\left[U^1[\theta], U^2[\theta]\right]=0\nonumber.\eea
The mapping of the LSP is straightforward. The wave functions are written $\Phi=\tau(\Psi)$ and satisfy
\bea\label{LSP'} D_1\Phi=\tau(u^1)\Phi=U^1[\theta]\Phi,\qquad 
        D_2\Phi=\tau(u^2)\Phi=U^2[\theta]\Phi.\eea
Again, for any generalized symmetry of the system of differential equations $\Omega[\theta]=0$ written in evolutionary form 
\be \label{wQ} \vec{w}_\Q=\Q^j[\theta]\frac{\partial}{\partial \theta^j}, \qquad \Q^j[\theta]\in \mathcal{B} \ee
there exists a surface $F\in \g$ whenever the tangent vectors given by 
\be \label{DF'} D_1F=\Phi^{-1} \left(pr\vec{w}_\Q U^1[\theta]\right) \Phi, \qquad  D_2F=\Phi^{-1}\left( pr\vec{w}_\Q U^2[\theta]\right) \Phi\ee
are linearly independent.   Here we use $\Q$ to denote the characteristic of the vector field $\vec{w}_\Q$ on jet space $N$ to distinguish it from the characteristic $Q$  of the vector field $\vec{v}_Q $ on jet space $M$. The following theorem holds as was first proven in \cite{GP2011a}.

\begin{theorem}
Suppose that the smooth potential matrices $U^\alpha[\theta]\in \g$ satisfy \eref{Delta'}  and $\Phi\in G$ satisfies the LSP \eref{LSP'}. Suppose further that there exists some generalized vector field $\vec{w}_\Q$ which is a symmetry of \eref{Delta'}. Then, there exists a $\g$-valued function $F$ whose tangent vectors satisfy \eref{DF'}, which can be written, up to a $\g$-valued constant, by 
\be\label{F'} F=\Phi^{-1} pr\vec{w}_\Q \Phi\ee
if and only if $\vec{w}_\Q$ is a symmetry of the LSP \eref{LSP'} in the sense that, for $\alpha=1,2$,
\be\label{deteqLSP'} pr\vec{w}_\Q\left(D_\alpha\Phi-U^\alpha[\theta]\Phi\right)=0, \qquad \mbox{ whenever } D_\alpha\Phi-U^\alpha[\theta]\Phi=0.\ee
\end{theorem}
To prove this theorem, one need only check that the compatibility conditions for \eref{DF'} are equivalent to the determining equations for a symmetry of \eref{Delta'}. Similarly, it is a direct computation to  show that the tangent vectors of  $F$ given by \eref{F'}  coincide with those given by  \eref{DF'} if and only if \eref{deteqLSP'} holds. 

We can consider the parameterization of the matrix functions $U^\alpha[\theta]$  as the Lax pairs for some integrable differential equations. Below, we distinguish three possible choices of parameterizations of potential matrices $U^\alpha[\theta]\in \g$:
\begin{itemize}
\item[ 1.]{\bf Zero-curvature representation for PDEs} In this case, suppose that the matrices $U^\alpha$ depend on both the independent variables $\xi_i=x_i$ $i=1,2$ and some dependent variables $f^j(x_1, x_2)$, as well as a spectral parameter $\lambda$ such that the ZCC $\Delta[u]=0$ is equivalent to a system of PDEs independent of the spectral parameter (see e.g.  \cite{ZakSha1974, ZakSha1978}) 
\bea \tau(\Delta[u])&=&D_2U^1([f],\lambda)-D_1U^2([f],\lambda)+\left[U^1([f],\lambda),U^2([f],\lambda)\right]\nn
&=&\Omega[f]=0.\label{Delta'f} \eea
This was the case treated in \cite{GP2011a} where the surface considered were associated with symmetries of the ZCC $\Delta[u]=0$ and the  PDE \eref{Delta'f}. 
\item[2.] {\bf Lax form for ODEs} Suppose now that we have only one independent variable $\xi_1=x_1$ and several dependent functions of one variable $g^j(x_1)$. The matrices $U^\alpha$ are functions on the jet space defined by $x_1$ and $g^j(x_1)$ and depend also on some spectral parameter $\lambda$ but not on $\xi_2=x_2$. The ZCC $\Delta[u]=0$ is thus equivalent to a system of ODEs which is assumed to be independent of the spectral parameter (see e.g.  \cite{Lax1968})
\bea \tau(\Delta[u])&=&-D_1U^2([g],\lambda)+\left[U^1([g],\lambda),U^2([g],\lambda)\right]\nn
&=&\Omega[g]=0.\label{Delta'g} \eea
This is the case treated in \cite{GP2011b}. Here again, the surfaces considered were associated with symmetries of the ZCC $\Delta[u]=0$ and the  ODE \eref{Delta'g}.
\item[3.] {\bf Zero-curvature representation for ODEs} Suppose now that we have several dependent functions $x^j(t)$ which depend only on one independent variable $\xi_1=t$. The matrices $U^\alpha$ are functions on the jet space defined by $t$ and $x^j(t)$ and the other independent variable, which here takes the form of a spectral parameter $\xi_2=\lambda.$ In this case, the ZCC $\Delta[u]=0$ is equivalent to a system of ODEs which is assumed to be independent of the spectral parameter (see e.g. \cite{BobEitbook})
\bea \tau(\Delta[u])&=&D_\lambda U^1([x],\lambda)-D_tU^2([x],\lambda)+\left[U^1([x],\lambda),U^2([x],\lambda)\right]\nn
&=&\Omega[x]=0,\label{Delta'x}\eea
where $D_t$ and $D_\lambda$ represent the total derivatives with respect to $t$ and $\lambda$ respectively.
This is the situation to be treated for the remainder of the paper and th surfaces are defined via symmetries of the ZCC $\Delta[u]=0$ and the ODE \eref{Delta'x}.
\end{itemize}

Note that in the first two cases, the spectral parameter, $\lambda,$ enters as a parameter in the LSP \eref{LSP'} whereas, in the third case, it enters as an independent variable in the LSP. Thus, the Sym-Tafel (ST) formula (see e.g \cite{Dodd1998, Kono1996, Sym, Tafel}) has a different realization in these cases. In the first two cases the independent variables are $x_1$ and $x_2$ and the ST formula, give by 
\be F^{ST}(x_1, x_2)=s(\lambda) \Phi^{-1} D_\lambda \Phi \label{STx}, \ee
is associated with a conformal symmetry in the spectral parameter of the equations \eref{Delta'f} and \eref{Delta'g} respectively. The tangent vectors of the ST formula in these cases are 
\bea D_1F^{ST}(x_1, x_2)=s(\lambda) \Phi^{-1}D_\lambda U^1\Phi,\qquad D_2F^{ST}(x_1, x_2)=s(\lambda) \Phi^{-1}D_\lambda U^2\Phi\nonumber .\eea
On the other hand,  in the third case, the independent variables are $t$ and $\lambda$ and  the immersion function (in analogy with \eref{STx}) 
\be F^{ST}(t, \lambda)=s(\lambda) \Phi^{-1} D_\lambda \Phi \label{Stl}, \ee is associated with a symmetry of the ZCC condition \eref{Delta}, which is a linear combination of a conformal symmetry in the spectral parameter $\xi_2=\lambda$ and a scaling of the potential matrices $u^1$ and $u^2$. The vector field $\vec{v}_Q$, associated with this symmetry, has characteristics given by \eref{vecST} and the tangent vectors of the surface are 
\bea  D_t F^{ST}(t,\lambda)=s(\lambda)\Phi^{-1}D_\lambda U^1 \Phi,\qquad \nonumber
D_\lambda F^{ST}(t,\lambda)=\Phi^{-1}D_\lambda\left(s(\lambda)U^2\right) \Phi
.\eea

\section{Lax pairs for an ODE written in zero-curvature form}
In this section, we shall consider specifically case 3 above; that is, ODEs with independent variable $t$ and dependent variable $x(t)$ which admit Lax pairs $U^\alpha([x],\lambda)$ that satisfy 
\be D_\lambda U^1([x],\lambda)-D_tU^2([x],\lambda)+\left[U^1([x],\lambda),U^2([x],\lambda)\right]=0.\ee
This type of Lax pair has been considered for Painlev\'e equations by several authors including \cite{ConteMusette, FIKNbook}.  Particularly relevant is \cite{BobEitbook} where the LSP was used to represent  the Painlev\'e equations and derive  the geometry of surfaces associated with a group-valued wave function  $\Phi$ \eref{LSP} with compatibility conditions \eref{Delta}. On the other hand,  the analysis contained herein  is  concerned  with surfaces given by  $\g$-valued functions F with compatibility conditions \eref{deteq} that  are infinitesimal deformations of \eref{Delta}. 
 
 The surfaces to be considered are given by $F([x],\lambda)\in \g$ with tangent vectors
 \bea\label{FAB}  D_t F=\Phi^{-1} A([x],\lambda) \Phi, \qquad  D_\lambda F=\Phi^{-1} B([x],\lambda) \Phi,\eea
 where the $\g$-valued matrices $A$ and $B$ satisfy 
 \be\label{AB'} D_\lambda A-D_t B+[A,U^2]+[U^1,B]=0.\ee
 Recall that the group-valued function $\Phi=\tau(\Psi)$ satisfies 
 \be\label{LSP'lt} D_t\Phi=U^1\Phi , \qquad  D_\lambda\Phi=U^2\Phi.\ee
 The possible forms of $A$ and $B$ can be decomposed into symmetries $\vec{v}_Q$ of the ZCC $\Delta[u]=0$ and $\vec{w}_\Q$ of the integrable equation $\Omega[x]=0$.  They are 
 \bea A([x],\lambda)=\tau\left(pr\vec{v}_Q(u^1)\right)+pr\vec{w}_\Q(U^1),\\
         B([x],\lambda)=\tau\left(pr\vec{v}_Q(u^2)\right)+pr\vec{w}_\Q(U^2).\eea
 Note that these two classes of symmetries are different. For example, the ZCC admits translation symmetries with respect to the independent variables whereas the integrable equation it is equivalent to may not admit such symmetries. This will be the case for the Painlev\'e equations considered in the next section. 
 
Using a linear combination of the symmetries in subsection \ref{symalpha}, the tangent vectors can be written 
 \bea \fl A([x],\lambda)=\alpha_1D_tU^1+\alpha_2D_\lambda U^1+\alpha_3 \left(t D_t U^1+U^1\right)+\alpha_4 \lambda D_\lambda U^1+ \alpha_5\left(D_t^2U^1+[D_tU^1,U^1]\right)\nn
 +\alpha_6 pr\vec{w}_\Q U^1\label{Aalpha}\eea
 and \bea
\fl\label{Balpha} B([x],\lambda)=\alpha_1D_tU^2 +\alpha_2 D_\lambda U^2+\alpha_3 t D_t U^2+\alpha_4 \left( \lambda D_\lambda U^2+U^2\right)+\alpha_5\left(D_t^2U^2+[D_tU^2,U^1]\right)\nn
+\alpha_6 pr\vec{w}_\Q U^2, \eea
where the $\alpha_i$'s are constants. 
We define the Fokas-Gel'fand formula for immersion in Lie algebras to be the immersion function $F$ with tangent vectors given by \eref{FAB} which are constructed out of a linear combination of symmetries $\vec{v}_Q$ of $\Delta[u]=0$ and $\vec{w}_\Q$ of $\Omega[x]=0$, as in \eref{Aalpha} and \eref{Balpha}. In what follows we will refer to it as such. This formula contains a linear combination of the immersion functions defined by \eref{F} and \eref{F'}.
 The integrated form of the surface with tangents defined by \eref{Aalpha} and \eref{Balpha}, is given by 
 \be\label{Falpha} \fl F=\phi^{-1}\left(\alpha_2D_\lambda +\alpha_1D_t +\alpha_3t U^1+\alpha_4 \lambda U^2+\alpha_5D_tU^1 +\alpha_6 pr\vec{w}_\Q \right)\Phi,\ee
 if and only if $\vec{w}_\Q$ is a generalized symmetry of the LSP \eref{LSP'}, in the sense of Theorem 2. 

 Note here that, on the one hand,  there exist exact integrated forms of $F$ for symmetries parameterized by the constants $\alpha_i$ for $i=1,\ldots,5.$ However, in order to check if the immersion function $F$ given by \eref{Falpha} has the form 
\be F^{i}=\tau(\Psi^{-1}pr\vec{v}_{Q_i}\Psi),\ee it is required to have a general solution for the wave function $\Psi$ so as to verify whether $\vec{v}_Q$ is a symmetry of the LSP \eref{LSP}. For example,  only with an exact form of the wave function $\Psi$ is it possible to say whether the LSP in invariant under translations in $t$. On the other hand, an integrated form of  the surface $F^{6}$ can be given only in the case that $\vec{w}_\Q$ is a symmetry of $\Omega[\theta]=0$ and its transformed LSP \eref{LSP'}. In this case, the integrated form is 
 \be \label{F6} F^{6}=\Phi^{-1}pr\vec{w}_\Q(\Phi).\ee

\section{Soliton surfaces via Painlev\'e equations P1, P2, and P3}
 
Next, we present some examples which illustrate the theoretical considerations described in sections 2-4. Various cases of the Painlev\'e type P1, P2, and P3  equations have been chosen in order to construct associated soliton surfaces whose GMC equations are equivalent to infinitesimal deformations of the Painlev\'e equations. 

\subsection{On the application of the method}
Here we shall briefly describe the present state of the analytic approach for finding soliton surfaces through the Fokas-Gel'fand formula for immersion associated with differential equations. In particular, we focus on difficulties arising in the attempt to explicitly construct  surfaces immersed in the Lie algebra. The basic method for solving this problem requires three parts for an explicit representation of the immersion function $F$ (given by \eref{Falpha}), namely
\begin{enumerate}
\item[i)] A zero-curvature representation of the ODE $\Omega[x]=0.$
\item[ii)] A generalized symmetry, $\vec{w}_\Q$, of the ODE $\Omega[x]=0.$
\item[iii)] A solution $\Phi$ of the LSP \eref{LSP'}.
\end{enumerate}
Note here that (i) is always required. However, even without the remaining two conditions some analysis of the induced immersion function $F$ \eref{Falpha} can be performed. For example, without an explicit form of the generalized symmetry it is still possible to  consider surfaces associated with terms $\alpha_i$ for $i=1\ldots 5$ in an integrated form \eref{Falpha}. Also, the geometry associated with the sixth term in \eref{Falpha} can be computed whenever $\Q$ satisfies the determining equation for a generalized symmetry of the considered model. 

In the third case, if we do not have an explicit form of the wave function $\Phi$ it is still possible to consider the geometric characteristics of the surfaces using the Killing form. Since the Killing form is invariant under group conjugation, the metric  or pseudo-metric on the tangent vectors \eref{FAB}, with \eref{Aalpha} and \eref{Balpha}, as well as the normals, defined up to a normalization factor by
\be N=\Phi^{-1}\left[A,B\right]\Phi,\ee
 are independent of the wave function $\Phi.$
The Killing form \cite{Helgason} is a symmetric bilinear product $\mathcal{B}(X,Y)$, given (up to a normalization factor) by
\be \label{B} \mathcal{B}(X, Y) =\frac12 tr(X\cdot Y),\qquad  X,Y\in \g.\ee
In terms of the basis $\{e_1, e_2, e_3\}$ for $sl(2,\mathbb{R})$, 
\be\label{basis} e_1=\left[\ba{cc}1&0\\ 0&-1 \ea \right], \quad e_2=\left[\ba{cc}0&1\\ 1&0 \ea \right], \quad e_3=\left[\ba{cc}0&-1\\ 1&0 \ea \right],\ee
the matrices $X$ and $Y$ can be decomposed as $X=X^je_j $ and $Y=Y^je_j$ for $j=1,2,3$ and the scalar product $\mathcal{B}$ defines a pseudo-Euclidean metric as 
\be \mathcal{B}(X,Y)=X^j\mathcal{B}_{jk}Y^k, \qquad \mathcal{B}_{jk}= \left[\ba{ccc}1 &0 &0\\
0 & 1& 0 \\ 0&0&-1\ea \right].\ee
 On $sl(2,\mathbb{R})$, the Killing form has signature $(2,1)$ and so induces a pseudo-Euclidean metric on the tangent vectors to the 2D-surface given by the immersion function $F\in sl(2, \mathbb{R})$. With this inner product, the surfaces are pseudo-Riemannian manifolds \cite{Eisenhart, doCarmo}. 
 
 In what follows, the surfaces considered are defined by a single smooth immersion function $F$ as the independent variables range over some subset $(t, \lambda)\in \Sigma \subset \mathbb{C} \cup \{ \infty\}.$ The singularity structures of the Painlev\'e equations are transported onto the surfaces and, in particular, the surfaces may have unbounded components.

\subsection{Painlev\'e P1}
In this section, we give several examples of surfaces associated with the Lax pair for the first Painlev\'e equation, P1,
\be \label{P1} x_{tt}-6x^2-t=0\ee
An LSP for the  Painlev\'e equation P1 \eref{P1} 
is given in terms of the potential matrices $U^1( [x],\lambda )$ and $U^2( [x],\lambda)$ taking values in the Lie algebra $ sl(2,\mathbb{R})$ \cite{JM1981, Kit1994}
\bea \label{U1P1} U^1=\left[\ba{cc} 0 &\lambda+2x\\ 1 & 0\ea\right], \\
	 \label{U2P1} U^2=\left[\ba{cc} -x_t &2\lambda^2+2\lambda x+t+2x^2\\ 2(\lambda-x) & x_t\ea \right],\nonumber\eea
which satisfy the zero-curvature condition
\be\label{ZCCP1} D_\lambda U^1-D_tU^2+[U^1, U^2]=(x_{tt}-6x^2-t)e_1 \nonumber.\ee
 Thus, there exists a wave function $\Phi$ of the LSP \eref{LSP'lt} taking values in the group $SL(2,\mathbb{R})$ if and only if $x(t)$ is a solution of the P1 equation \eref{P1}. 
	
Let us  consider the surface associated with translation in the variable $t$
	 \[ F^{1}= \Phi^{-1}D_t \Phi.\]
The tangent vectors to the surface are given by \eref{FAB} with 	 
\bea A=D_tU^1=\left[ \begin {array}{cc} 0&2\,x_{{t}}\\\noalign{\medskip}0&0
\end {array} \right]\in sl(2,\mathbb{R}), \nn 
B=D_tU^2=\left[ \begin {array}{cc} -6\,{x}^{2}-t&1+2\lambda x_{{t}}+4xx_{{t}
}\\\noalign{\medskip}-2\,x_{{t}}&6\,{x}^{2}+t\end {array} \right]\in sl(2,\mathbb{R}).\nonumber\eea
The first fundamental form for the surface is 
 \bea I(F^{1})=-4{x_{{t}}}^{2}dtd\lambda+\left(36\,{x}^{4}+12\,{x}^{2}t+{t}^{2}-2\,x_{{t}}-4\,{x_{{t}}}^{2}\lambda-8
\,{x_{{t}}}^{2}x\right)d\lambda^2.\nonumber\eea
Note that the tangent vector  $D_tF^{1}$ is an isotropic vector. 
The normal to the surface $F^{1}$ is 
\be N=\Phi^{-1}\left[ \begin {array}{cc} 1&-{\frac {6\,{x}^{2}+t}{x_{{t}}}}
\\\noalign{\medskip}0&-1\end {array} \right]\Phi\in sl(2,\mathbb{R}),\nonumber\ee
which allows for the computation of the second fundamental form and the Gaussian and mean curvatures
\bea II(F^{1})=2x_{{t}}dt^2+8x_{{t}} \left( \lambda-x \right)dtd\lambda\nn
\qquad +2{\frac { \left( \lambda-x \right)\left( 4{x_{{t}}}^{2}\lambda+8
{x_{{t}}}^{2}x-36{x}^{4}+x_{{t}}-{t}^{2}-12\,{x}^{2}t \right) }{x_
{{t}}}}d\lambda^2,\nn
K(F^{1})={\frac { \left( -12\,{x_{{t}}}^{2}x-x_{{t}}+36\,{x}^{4}+12\,{x}^{2}t+{
t}^{2} \right)  \left( \lambda-x \right) }{{x_{{t}}}^{4}}},\nn
H(F^{1})= -{\frac {36\,{x}^{4}+12\,{x}^{2}t+{t}^{2}-2\,x_{{t}}+4\,{x_{{t}}}
^{2}\lambda-16\,{x_{{t}}}^{2}x}{4{x_{{t}}}^{3}}}.\nonumber\eea
Note that all the points on the line $\lambda=x(t)$, where $x(t)$ is a solution of the Painlev\'e equation P1, are parabolic points ($K=0$). 
The  umbilical points, where the principal curvatures coincide, on the surface satisfy
\be H^2-K=c_2x_t^{-2}+c_3x_t^{-3} +c_4x_t^{-4} +c_5x_t^{-5} +c_6x_t^{-6}=0,\ee
where
\bea
c_2=\left( \lambda+2\,x \right) ^{2}, \qquad c_3=3x,\nn
c_4=-\left(x+\frac{\lambda}{2}\right) (6x^2+t)^2+\frac14, \nn
c_5=-\frac14 (6x^2+t)^2,\qquad 
c_6=\frac1{16}(6x^2+t)^4\nonumber.\eea
It is straightforward to observe that the umbilical points on the surface $F^1$ lie on the curves
\[ \lambda=-2x_t+\frac{(6x^2+t)^2}{4x_t^2}\pm \frac{1}{x_t^2}\sqrt{x_t(6x^2+t)^2-12x_t^2(xx_t-1)},\]
whenever $x_t\ne 0.$  

 Next we consider  the surfaces $F^{2}$ associated with translation in the spectral parameter $\lambda$, 
 	 \[ F^{2}=\Phi^{-1}D_\lambda \Phi.\]
 The tangent vectors to the surface are defined as in  \eref{FAB} with 
\bea A=D_\lambda U^1= \left[ \begin {array}{cc} 0&1\\\noalign{\medskip}0&0\end {array}
 \right]\in sl(2,\mathbb{R}) , \nn
  B=D_\lambda U^2= \left[ \begin {array}{cc} 0&4\,\lambda+2\,x\\\noalign{\medskip}2&0
\end {array} \right]\in sl(2,\mathbb{R}).\nonumber \eea
The first fundamental form associated with this surface is given by 
\be I(F^{2})=2dtd\lambda+4(x+2\lambda)d\lambda^2.\nonumber\ee
Again, the tangent vector $D_tF^{2}$ is an isotropic vector. 
The normal to the surface is 
\be
 N=\Phi^{-1}e_1\Phi\in sl(2,\mathbb{R}).\nonumber \ee
Thus, the image of the surfaces $F^{2},$ written in the moving frame defined by the (nonconstant) wave function $\Phi$,  lies in a plane. The second fundamental form and Gaussian and mean curvature for this surface are 
\bea II(F^{2})=-dt^2+4(x-\lambda)dtd\lambda+2\left(4x^2+4\lambda x+t-\lambda^2\right)d\lambda^2,\\
K(F^{2})=2(6x^2+t)=2x_{tt},\\
H(F^{2})=2(2x+\lambda).\eea
Note that the Gaussian curvature does not depend on the spectral parameter $\lambda$ and in fact the sign of the second derivative of the solution $x(t)$ of the Painlev\'e equation P1 \eref{P1} determines whether the points of the surface are hyperbolic, elliptic or parabolic. Umbilical points of the surface are determined by 
\[ H^2-K=4(2x+\lambda)^2-2(6x^2+t)=0,\]
which are exactly the curves 
\[ \lambda=-2x\pm \sqrt{\frac{6x^2+t}{2}}=-2x\pm \sqrt{\frac{x_{tt}}{2}}.\]
There are no umbilical points in the regimes where $x_{tt}<0.$

Let us now consider the surfaces $F^{5}$ associated with a generalized symmetry of the ZCC, $\Delta[u]=0$. 
The surface $F^{5}$ is given by
\bea F^{5}=\Phi^{-1}D_tU^1\Phi,\eea
with tangent vectors
\bea  D_t F^{5}=\Phi^{-1}\left(D_t^2U^1+[D_tU^1,U^1]\right)\Phi,\nn \nonumber
D_\lambda F^{5}=\Phi^{-1}\left(D_t^2U^2+[D_tU^2,U^1]\right)\Phi.\eea 
For the particular choices of potential matrices $u^\alpha=U^\alpha([x],\lambda)$ as in \eref{U1P1} and \eref{U2P1}, the tangents vectors are given as in \eref{FAB}
with 
\bea A=\left[ \begin {array}{cc} 2\,x_{{t}}&12\,{x}^{2}+2\,t
\\\noalign{\medskip}0&-2\,x_{{t}}\end {array} \right]\in sl(2,\mathbb{R}), \nonumber\\
B= \left[ \begin {array}{cc} -4\,xx_{{t}}+4\,x_{{t}}\lambda&4\,{x_{{t}}}
^{2}\\\noalign{\medskip}0&4\,xx_{{t}}-4\,x_{{t}}\lambda\end {array}
 \right]\in sl(2,\mathbb{R}).\nonumber\eea
 Although the tangent vectors are linearly independent, the first fundamental form on the surface  
 \be I(F^{5})=4x_t^2\left(dt^2-2(x-\lambda)dtd\lambda +4(x-\lambda)^2d\lambda^2\right),
\nonumber\ee
 is degenerate ($det(g_{ij})=0$) since the normal to the surface is an isotropic vector. Thus the surface given by the immersion function $F^{5}$ lies in a plane in the moving frame defined by conjugation by $\Phi.$ 

Let us now use a generalized symmetry of the ODE \eref{P1} to induce the surface. That is, suppose  that there exists a generalized vector field in evolutionary representation
 \be  \vec{w}_\Q =\Q[x]\frac{\partial }{\partial x}\ee
 which is a symmetry of P1. The determining equation for $\Q[x]$ is obtained from the second prolongation of the vector field $\vec{w}_\Q$  applied to the P1 equation
 \be \label{detQP1} D_t^2\Q-12 x\Q=0, \qquad \mbox{ whenever } x_{tt}-6x^2-t=0.\ee
Expanding the total derivative $D_t$  in terms of partial derivatives with respect to $t, x, x_t$, taken whenever  P1 \eref{P1} holds, gives
 \bea\fl D_t\Q&=& \frac{\partial }{\partial t} \Q+x_t\frac{\partial}{\partial x}\Q +(6x^2+t)\frac{\partial }{\partial x_t}\Q,\nonumber\\
\fl D_t^2\Q&=&  \frac{\partial ^2}{\partial t^2}\Q +2x_t\frac{\partial^2}{\partial t \partial x}\Q+2(6x^2+t)\frac{\partial^2}{\partial t\partial  x_t}\Q+x_t^2\frac{\partial ^2}{\partial x^2}\Q +2x_t(6x^2+t)\frac{\partial ^2}{\partial x \partial x_t}\Q\nn \fl
 &&+(6x^2+t)^2\frac{\partial ^2}{\partial x_t^2}\Q+(6x^2+t)\frac{\partial }{\partial x}\Q+(12xx_t+1)\frac{\partial}{\partial x_t}\Q\nonumber.\eea
Thus, the determining equation \eref{detQP1} becomes a linear, analytic second-order PDE for $\Q$
\bea\fl   \frac{\partial ^2}{\partial t^2}\Q +2x_t\frac{\partial^2}{\partial t \partial x}\Q+2(6x^2+t)\frac{\partial^2}{\partial t x_t}\Q+x_t^2\frac{\partial ^2}{\partial x^2}\Q +2x_t(6x^2+t)\frac{\partial ^2}{\partial x \partial x_t}\Q\label{detQP1exp}\\ \!\!\!\!\!\!\!\!\!\!\!\! +(6x^2+t)^2\frac{\partial ^2}{\partial x_t^2}\Q+(6x^2+t)\frac{\partial }{\partial x}\Q+(12xx_t+1)\frac{\partial ^2}{\partial x_t}\Q-12x\Q=0 \nonumber .\eea
The construction of generalized symmetries of the Painlev\'e equation P1 requires a solution for $\Q$ as a function of $t, x, x_t$. For equation \eref{detQP1exp}, the Cauchy-Kovalevskaya theorem ensures that analytic solutions exist provided that the initial data is analytic (see e.g. \cite{FJon, Olver}). So, in what follows, we assume that the quantity $\Q$ is a given function of $t, x,$ and $ x_t$ satisfying \eref{detQP1exp}. 

As proven in section \ref{integrable}, if the vector field $\vec{w}_\Q$ is a symmetry of the P1 equation \eref{P1} then the matrices 
\bea A=pr\vec{w}_\Q(U^1)= \left[ \begin {array}{cc} 0&2\,\Q\\\noalign{\medskip}0&0\end {array}
 \right]\in sl(2,\mathbb{R}), \nonumber\\
  B=pr\vec{w}_\Q(U^2)=\left[ \begin {array}{cc} -D_{{t}} \Q &\Q \left( 2\,
\lambda+4\,x \right) \\\noalign{\medskip}-2\,\Q&D_{{t}} \Q
  \end {array} \right]\in sl(2,\mathbb{R}), \nonumber\eea
satisfy \eref{AB'} and hence there exists a surface $F^6$ with tangent vectors given by \eref{FAB}. 
The first fundamental form of this surface is
\bea I(F^{6})=-4\,{\Q}^{2}dtd\lambda+\left(  \left( D_{{t}}\Q   \right) ^{2}-4\,{\Q}^{2}\lambda-8\,{\Q
}^{2}x\right)d\lambda^2,\nonumber\eea
	Note that, in the surface $F^{6}$ the tangent vector in the direction $D_t$ is again an isotropic vector.
The normal vector
\bea N=\Phi^{-1}\left[ \begin {array}{cc} 1&-D_t\left(\ln (\Q)\right)
\\\noalign{\medskip}0&-1\end {array} \right]\Phi,\nonumber\eea
allows for the construction of the second fundamental form and Gaussian and mean curvatures
\bea \fl II(F^{6})=2 \Q dt^2+8 \Q \left( \lambda-x \right) dtd\lambda\nn
\fl \qquad  +\frac{2}{\Q}\left( 4{\Q}^{2}{\lambda}^{2}+4 {\Q}^{2}\lambda x+{\Q}^{2}t-2{\Q
}^{2}{x}^{2}-(D_{{t}}  \Q ) x_{{t}}\Q- (x+\lambda)\left(  D_{{t}} 
\Q \right)  \right) ^{2}d\lambda^2,\nonumber\\ 
\fl 
K(F^{6})={\frac{-1}{\Q^{4}}}\left({12\,{\Q}^{2}\lambda\,x+{\Q}^{2}t-6{\Q}^{2}{x}^{2}- \left( D_{{t}}
\Q \right) x_{{t}}\Q- \left( D_{{t}} \Q   \right) ^
{2}\lambda+ \left( D_{{t}} \Q \right) ^{2}x}\right),\nn \fl 
H(F^{6})=\frac{-1}{4\Q^3} \left(\left( D_{{t}}  \Q   \right) ^{2}+4\,{\Q}^{2
}\lambda-16\,{\Q}^{2}x \right)\nonumber.\eea
 The mean and Gaussian curvatures depend on the Painlev\'e trancendent P1 and on the solution of its associated determining equation \eref{detQP1exp}.
The umbilical points of the surface $F^6$ satisfy
\[ H^2-K=c_{4}D_t^4\Q+c_2D_t^2\Q+c_1D_t\Q +c_0=0,\]
where 
\bea c_4=\frac{1}{16\Q^6}, \qquad c_2=-\frac{2x+\lambda}{2\Q^4},\nn
c_1=\frac{-x_t}{\Q^3}, \qquad c_0=\frac{\lambda^2+4\lambda x+10x^2+t}{\Q^2}\nonumber,\eea
the solution of which lie on the curves on the surface $F^6$
\[ \lambda=\frac{1}{4}\left(D_t(\ln \Q)\right)^2-8x\pm \sqrt{x_t D_t(\ln \Q)-x_{tt}}\]

\subsection{Painlev\'e P2}
In this section, we give several examples of surfaces associated with the Lax pair for the second Painlev\'e equation P2
 \be \label{P2} x_{tt}=2x^3+tx-\alpha.\ee
A Lax pair for this equation is given by  \cite{Garn1960, ItsKap1988}
 \bea U^{1}= \left[ \begin {array}{cc} -\lambda&x\\\noalign{\medskip}x&\lambda
\end {array} \right] 
\in sl(2,\mathbb{R}),\nonumber\\
U^{2}= \left[ \begin {array}{cc} 4\,{\lambda}^{2}-2\,{x}^{2}-t&-4\,x\lambda+
{\frac {\alpha}{\lambda}}+2\,x_{{t}}\\\noalign{\medskip}-4\,x\lambda+{
\frac {\alpha}{\lambda}}-2\,x_{{t}}&-4\,{\lambda}^{2}+2\,{x}^{2}+t
\end {array} \right] 
 \in sl(2, \mathbb{R}) ,\nonumber
 \eea

 First, we consider surfaces $F^{1}$ associated with translation in the $t$-direction,
 	 \be F^{1}=\Phi^{-1}D_t \Phi\in sl(2, \mathbb{R}),.\ee
 	 %
 The tangent vectors to the surface are defined as in  \eref{FAB} 
 \bea D_tF^{1}=\Phi^{-1}D_t U^{1}\Phi, \nonumber \\
 		 D_\lambda F^{1}=\Phi^{-1} D_t U^{2}\Phi^{-1},\nonumber \eea
with matrices
\bea
A=D_t U^{1}=\left[ \begin {array}{cc} 0&ix_{{t}}\\\noalign{\medskip}-ix_{{t}}&0
\end {array} \right],  \nonumber \\
B=D_t U^{2} =\left[ \begin {array}{cc} -i \left( 1+4\,xx_{{t}} \right) &4\,ix_{{t}
}\lambda+4\,{x}^{3}-2\,{x}^{2}+2\,\alpha\\\noalign{\medskip}-4\,ix_{{t
}}\lambda+4\,{x}^{3}-2\,{x}^{2}+2\,\alpha&i \left( 1+4\,xx_{{t}}
 \right) \end {array} \right].\nonumber \eea
		%
The corresponding first fundamental form of the surface is given by 
\bea\fl\nonumber   I(F^{1})={x_{{t}}}^{2}dt^2 -8\lambda\,{x_{{t}}}^{2} dtd\lambda\\
+\left(16\left( {x}^{2}+{\lambda}^{2} \right) {x_{{t}}}^{2}+8\,xx_{{t}
}- 4\left( \,\alpha-2\,{x}^{3}-\,{x}^{2} \right)^2+1  \right)d\lambda^2,\nonumber\\
 \fl \nonumber det\left(g_{ij}(F^{1})\right)={x_{{t}}}^{2} \left((1+4xx_t)^2-4(\alpha-tx-2x^3)^2\right).  \eea
The (unnormalized) normal vector
\bea N(F^{1})&=&\Phi^{-1}\left[ \begin {array}{cc} -4\,{x}^{3}-2\,xt+2\,\alpha&1+4\,xx_{{t}}
\\\noalign{\medskip}-4\,xx_{{t}}-1&4\,{x}^{3}+2\,xt-2\,\alpha
\end {array} \right] \Phi
,\eea
%
				%
is  used to determine the second fundamental forms of the surface
\bea\fl II(F^{1})&=& -2\lambda \left( 1+4\,xx_{{t}} \right) x_{{t}}dt^2+4\,x_{{t}} \left( -4\,x_{{t}}\alpha-2\,{x}^{2}+4\,{\lambda}^{2}-t+16\,
xx_{{t}}{\lambda}^{2} \right) dtd\lambda\nn\fl &&
+\lambda^{-1}\Bigg[ 32\,{ { \left( {\lambda}^{2}+{x}^{2} \right)  \left( -4\,x{
\lambda}^{2}+\alpha \right) {x_{{t}}}^{2}}}+8\,{ {
 \left( -6\,{x}^{2}{\lambda}^{2}-4\,{\lambda}^{4}+{\lambda}^{2}t+2\,
\alpha\,x \right) x_{{t}}}}\nn \fl &&
-2\,{ { \left( 2\,\alpha+1-2
\,xt-4\,{x}^{3} \right)  \left( 2\,\alpha-1-2\,xt-4\,{x}^{3} \right) 
 \left( -4\,x{\lambda}^{2}+\alpha \right) }}\Bigg]d\lambda^2.\nonumber \eea
From the fundamental forms, it is possible to compute the Gaussian and mean curvatures.
The umbilical points on the surface satisfy a  polynomial equation in $\lambda,$ $x$ and $x_t.$ In each case, the formulas are too involved to be presented in an instructive manner.

 Consider next the surfaces associated with translation in the spectral parameter $\lambda$
 	 \be F^{2}=\Phi^{-1}D_\lambda \Phi\in sl(2, \mathbb{R}).\ee
 The tangent vectors to the surface are defined as in  \eref{FAB},  
 \bea D_tF^{2}=\Phi^{-1}D_\lambda U^{1}\Phi, \nn  D_\lambda F^{2}=\Phi^{-1}D_\lambda U^{2}\Phi\nonumber,\eea
 with matrices
 \bea A=  D_\lambda U^{1}=\left[ \begin {array}{cc} -1&0\\\noalign{\medskip}0&1\end {array}
 \right], \nonumber
 \\
B=D_\lambda U^{2} =\left[ \begin {array}{cc} 8\,\lambda&-4\,x-{{\alpha}{{\lambda}^
{-2}}}\\\noalign{\medskip}-4\,x-{{\alpha}{{\lambda}^{-2}}}&-8\,
\lambda\end {array} \right].\nonumber
 \eea
		%
The first fundamental form for this surfaces immersed in  $sl(2, \mathbb{R})$ is
\bea  I\left(F^{2}\right)=dt^2-16\lambda dtd\lambda+\left(\left(\frac{\alpha}{\lambda^2}+4x\right)^2+64\lambda^2\right)d\lambda^2,\nonumber\\
det\left(g_{ij}\left(F^{2}\right)\right)=\left(\frac{\alpha}{\lambda^2}+4x\right)^2.\label{gF2}
\eea
In the moving frame defined by conjugation by the (non-constant) wave function  $\Phi$, the normal to the surface is constant and so the images of the immersion function $F^{2}$ is contained in a plane.
The second fundamental form, Gaussian and mean curvatures for the surface in $sl(2, \mathbb{R})$ is given by 
 \bea II(F^{2})=2xdt^2-4(4x\lambda-\alpha\lambda^{-1})dtd\lambda\nn \qquad\qquad  +2\left(8x^3+2\alpha\lambda^{-2} x^2+(16\lambda^{2}+4t)x-\alpha\lambda^{-2}(12\lambda^2-t)\right)d\lambda^2,\nn
K(F^{2})=\frac{4\lambda^2 x_{tt}}{4x\lambda^2+\alpha},\nonumber\\
H(F^{2})=\frac{4\lambda^4+\lambda^2(6x^2+t)+\alpha x}{4x\lambda^2+\alpha}.\nonumber\eea

 Note here the sign of the second derivative of the solution $x(t)$ of the Painlev\'e equation P2 \eref{P1} as well as square root of the determinant of the metric \eref{gF2} determines whether the points of the surface are hyperbolic, elliptic or parabolic. The umbilical points of the surface $F^{2}$ satisfy 
\[\fl 16{\lambda}^{8}+ 8\left(6{x}^{2}+t \right) {\lambda}^{6}+
 \left( 4{x}^{4}-4t {x}^{2}+24x\alpha+{t}^{2} \right) {\lambda}^
{4}+2\alpha \left( 2{x}^{3}-tx+2\alpha \right) {\lambda}^{2}+{
x}^{2}{\alpha}^{2}=0.
\]

Next, consider surfaces associated with generalized symmetries of the Painlev\'e  P2 equation. 
Since we do not have explicit forms of the wave functions $\Phi$ which satisfy the LSP \eref{LSP'}, it is not possible to say whether the integrated form of the immersion function \eref{F6} holds. However, the geometry of the surfaces can be studied via the tangent vectors, defined as in  \eref{FAB},
 \bea\label{tanQP2a} D_tF^{6}=\Phi^{-1}\left(pr\vec{w}_\Q U^{1}\right)\Phi, \\
 			\label{tanQP2b}  D_\lambda F^{6}=\Phi^{-1}\left(pr\vec{w}_\Q U^{2}\right)\Phi.\eea
			%
Here $\Q$ is the characteristic of the vector field $\vec{w}_\Q$ written in evolutionary form \eref{wQ} and hence is required to satisfy the linear determining equation 
\be D_t^2\Q=6x^2\Q+t\Q,\qquad  \mbox{ whenever } x_{tt}=2x^3+tx-\alpha.\ee
The  matrices in \eref{tanQP2a} and \eref{tanQP2b} are given by 
\begin{eqnarray*} A=pr\vec{w}_\Q U^{1}=\left[ \begin {array}{cc} 0&\Q\\\noalign{\medskip}\Q&0\end {array}
 \right],
\\
 B= pr\vec{w}_\Q U^{2}= \left[ \begin {array}{cc} -4\,x\Q &-4\,\lambda\,\Q+2\,D_{{t}} \left( \Q
 \right) \\\noalign{\medskip}-4\,\lambda\,\Q-2\,D_{{t}} \left( \Q
 \right) &4\,x\Q\end {array} \right].
\end{eqnarray*}
 The first fundamental form and determinant of the induced scalar product are 
\bea I(F^{6})={\Q}^{2}dt^2-8\lambda \Q^2dtd\lambda -4\left((D_t\Q)^2-4(x^2+\lambda^2)\Q^2\right)d\lambda^2,\nonumber \\
det\left(g_{ij}(F^{6})\right)=-4\Q^2\left((D_t\Q)^2-4x^2\Q^2\right).\nonumber  \eea

The second fundamental forms and Gaussian and mean curvatures are 
\bea \fl II\left(F^{6}\right)=\frac{1}{\sqrt{(D_t\Q)^2-4x^2\Q^2}}\Bigg[4\lambda x\Q^2dt^2-8\,\Q \left( x_{{t}}D_{{t}}  \Q  +4\,x\Q{\lambda}^{2}-2{x}^{3}\Q-{x}^{2}\Q \right) dtd\lambda \nn
\fl \qquad +
 \frac1\lambda\Bigg(\left( -16\,x{\lambda}^{2}+4\,\alpha \right)  \left( D_{{t}} \Q  \right) ^{2}+16  x_{{t}}{\lambda}^{
2}\Q D_{{t}}\Q -16\,{\Q}^{2}x \left( -x{\lambda}^{2}+2\,{x}^{2}{\lambda}^{2}+\alpha
\,x-4\,{\lambda}^{4} \right) 
\Bigg)d\lambda^2\Bigg],\nn
\fl K\left(F^{6}\right)=\frac{-1}{ \left((D_t\Q)^2-4x^2\Q^2 \right) ^{2}}\Bigg( 4\left(- 4 {x}^{2}{\lambda}^{2}+\alpha x-{x_{{t}}}^{2}
 \right) \left( D_{{t}}\Q  \right) ^{2}+2x\Q x_{{t}}
 \left( 2{\lambda}^{2}-2{x}^{2}-x \right) D_{{t}} \Q\nn
 +{\Q}^{2}{x}^{3} \left( 4\,{x}^{3}+4\,{x}^{2}-24\,x{\lambda}^{
2}+x+4\,\alpha-4\,{\lambda}^{2} \right) \Bigg),\nn
\fl H\left(F^{6}\right)=\frac{1}{2\lambda \left(D_t(\Q)^2-4x^2\Q^2\right)^{\frac32}}\Bigg(\left( 8\,x{\lambda}^{2}-\alpha \right)  \left( D_{{t}} \Q  \right) ^{2}+4 x_{{t}}{\lambda}^{2
}\Q D_{{t}} \Q
-4{x}^{2}{\Q}^{2} \left( 10\,x{\lambda}^{2}+{\lambda}^{2}-\alpha
 \right) \Bigg).\nonumber\eea

In particular, we can use classical solutions of P2 to find explicit forms of the soliton surfaces associated with particular solutions. For example, it has been shown in \cite{Gromak1999}, that  Painlev\'e  P2 has rational solutions for integer values of $\alpha$. These rational solutions have recently emerged in studying the small dispersion or semi-classical limit of the KdV and sine-Gordon equations respectively \cite{BucMil2011, ClaGra2010}.
Below, we list the first two rational solutions of the Painlev\'e equation P2 (\ref{P2}) along with the characteristic of vector fields,  $\Q,$ 
which satisfy the determining equation
\be\label{detQP2} D_t^2\Q-(6x^2+t)\Q=0, \ee
for the given integer values of $\alpha$:
 \be\fl \ba{llll} &\alpha=1\qquad & x=t^{-1}  & \Q=\sqrt{t} I_\frac{5}3 (\frac{2t^\frac32}{3})\\
\fl& &\mbox{ or} & \Q=\sqrt{t} K_\frac{5}3 (\frac{2t^\frac32}{3})\\
\mbox{and} &\alpha=2 \qquad &x=-\frac{2(t^3-2)}{t(t^3+4)}  & \Q=\frac{1}{t^2(t^3+4)^2}HeunC(0,-\frac53,-5, \frac49,\frac{73}{18},-\frac{t^3}{4})\\
\fl &&\mbox{ or} & \Q=\frac{t^3}{(t^3+4)^2}HeunC(0,\frac53,-5, \frac49,\frac{73}{18},-\frac{t^3}{4})\ea \nonumber\ee
Here $I_\nu(z), K_\nu(z)$ are modified Bessel functions  \cite{Yellowbook} and $HeunC(\alpha, \beta, \gamma, \delta, \mu, t)$ is a solution of the confluent Heun equation \cite{Ronveaux}. 
The graphs of surfaces given by immersion functions $F^1$ and $F^2$ are displayed in figure 1 for the first two rational solutions of P2. 
\begin{figure}\label{gneg}
\begin{center}$
\begin{array}{cc}

\includegraphics[width=2.5in]{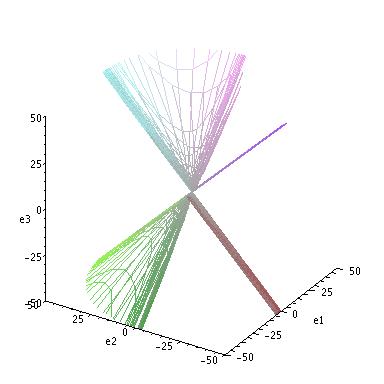}&
\includegraphics[width=2.5in]{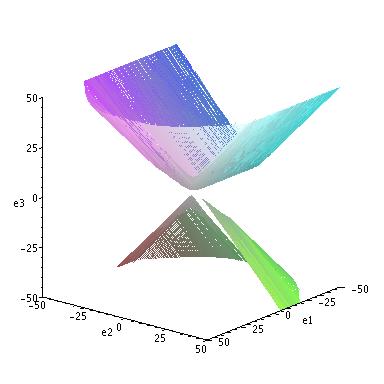} \\
F^{1}: \alpha=1 ,&F^{2}: \alpha=1,\\

\includegraphics[width=2.5in]{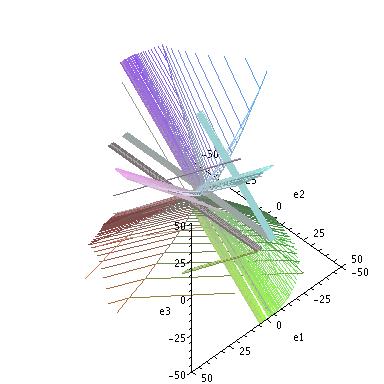}&
\includegraphics[width=2.5in]{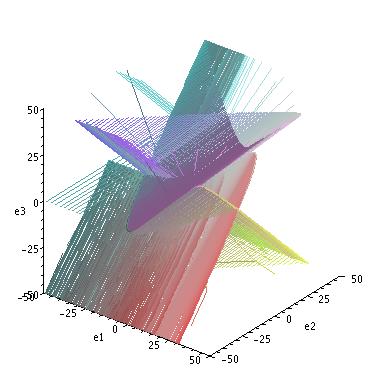} \\
F^{1}: \alpha=2 ,&F^{2}: \alpha=2,\end{array}$

\end{center}
\caption{Surfaces $F^{1}$ and $F^{2}$ for rational solutions of P2. The domain of parameterization is chosen to be $t\in [-30,30],$ $\lambda \in [-30,30]$ except for an omitted range of $\pm .01$ around the singularities $\lambda=0$ and $t=0$ for the first rational solution and $t=0,-2^{2/3}$ for the second rational solution. The viewing window for the graphs are $[-50,50]^3$ and the axes are the components of surface in the basis \eref{basis}.  }
\end{figure}

For the case $\alpha=\epsilon / 2$ with $\epsilon ^2=1$, a solution of P2 can be written in terms of the Airy function with \cite{AblSeg1977, Gromak1999}
\be x(t)=-\epsilon\frac{d}{dt}\ln\left(Ai\left(2^{-\frac13}t\right)\right),\qquad \epsilon^2=1,\ee
where $Ai(\cdot)$ is the Airy function. In this case, $x(t)$ satisfies the first-order differential equation 
\be \label{P212} x_t=\epsilon x^2+\epsilon \frac t2.\ee
The first fundamental forms and  normals for surfaces $F^1 $ and $F^6$, in the case $\alpha=\epsilon/2$,  are given by 
\bea I\left(F^{1}\right)=-(2x^2+t)^2\left(\frac{1}4dt^2-2\lambda dt d\lambda+4\lambda^2d\lambda^2\right),\\
 N\left(F^{1}\right)=\Phi^{-1}\left[\ba{cc} 1 &-1\\ 1 &-1\ea \right]\Phi,\eea
 and \bea 
I\left(F^{6}\right)=-\Q^2\left(dt^2-4\lambda dt d\lambda +4\lambda^2d\lambda^2\right),\\
 N\left(F^{6}\right)=\Phi^{-1}\left[\ba{cc} 1 &-1\\ 1 &-1\ea \right]\Phi,\eea
respectively. Here $\Q$ satisfies the determining equation for a symmetry of \eref{P212}, 
\bea \label{detP212} D_t\Q=\epsilon 2x\Q.\\
\Rightarrow\frac{D_t \Q}{\Q}=2\frac{d}{dt}\ln\left(Ai\left(2^{-\frac13}t\right)\right).\eea
The solution for this $\Q$ is given by 
\[ \Q =cAi(2^{-\frac13}t)^2. \]
Note that a solution $\Q$ of \eref{detP212} is also a solution of the general determining equation for a symmetry of P2 \eref{detQP2} whenever $x(t)$ is a solution of \eref{P212}. Indeed, 
\bea D_t^2\Q &=&D_t(\epsilon 2x\Q)=(6x^2+t)\Q .\eea

\begin{figure}\label{gneg}
\begin{center}$
\begin{array}{c}

\includegraphics[width=2.5in]{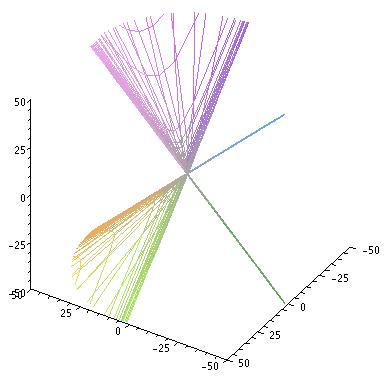},\end{array}$

\end{center}
\caption{Surface $F^{6}$ with $\alpha=0, \beta=1, \gamma=2/5, \delta=0$ and rational solution for $x(t).$ The domain of parameterization is chosen to be $t\in [-30,30],$ $\lambda \in [-30,30]$ except for an omitted range of $\pm .01$ around the singularities $\lambda=0$ and $t=0$. The viewing window for the graph is $[-50,50]^3$ and the axes are the components of surface in the basis \eref{basis}. }
\end{figure}
The surfaces $F^{1}$ and $F^{6}$, obtained from different symmetries,  are conformally equivalent and are contained in planes with isotropic normal vectors in the moving frame defined by conjugation by $\Phi$. Note that, with the restriction $\alpha=\epsilon/2,$ the ODE \eref{P212} and determining equation for $\Q$ \eref{detP212} are exactly those required so that the metrics induced by the Killing form on the tangents to the surfaces  $F^{1}$ and $F^{6}$ are degenerate, i.e. the determinants of the first fundamental forms are zero. As above,  the tangent vectors to the surface are linearly independent and so the immersion functions define surfaces instead of a curves.   A graph of the surface $F^6$ for the symmetry $\vec{w}_{R_1}$ is given in figure 2. 
 
 \subsection{Painlev\'e  P3} 
 In this section, we consider the cases of the third Painlev\'e equation P3 which admit point symmetries and their associated surfaces. In these cases, we use the known solutions of the determining equations for the vector field to obtain explicit expressions for the geometry of such surfaces.

 The third Painlev\'e equation, P3, is given by
 \be \label{P3} x_{tt}=\frac{(x_t)^2}{x}-\frac{x_t}{t}+\frac1t(\alpha x^2+\beta)+\gamma x^3+\frac{\delta}{x}.\ee
The LSP is given in terms of the potential matrices $U^1(\lambda, [x]) $ and $U^2(\lambda, [x])$ taking values in the Lie algebra $sl(2, \mathbb{R})$ \cite{ConteMusette, Garn1960, Gromak1999} 
 \bea\fl  U^1(\lambda, [x])=\frac12 \left[ \begin {array}{cc} \,{\frac {x_{{t}}}{x}}+\,\gamma\,x+\,{\frac {\delta}{x}}&2\lambda\\\noalign{\medskip}2\lambda&-\,{
\frac {x_{{t}}}{x}}-\,\gamma\,x-\,{\frac {\delta}{x}}
\end {array} \right],\\
\fl  U^2(\lambda, [x])= \left[ \begin {array}{cc} {\frac {2\,t{\lambda}^{2}x_{{t}}+2\,t{
\lambda}^{2}\gamma\,{x}^{2}+2\,t{\lambda}^{2}\delta-x\alpha\,\delta+x
\beta\,\gamma}{4\lambda\,x \left( {\lambda}^{2}+\gamma\,\delta \right) 
}}
&-{\frac {-2\,t{\lambda}^{2}+\gamma\,tx_{{t}}+\gamma\,x+{\gamma
}^{2}t{x}^{2}-\gamma\,\delta\,t+\alpha\,x}{2({\lambda}^{2}+\gamma\,
\delta)}}\\
\noalign{\medskip}-{\frac {-2\,t{x}^{2}{\lambda}^{2}+
\delta\,tx_{{t}}-\delta\,x+{\delta}^{2}t-\gamma\,\delta\,t{x}^{2}-
\beta\,x}{2{x}^{2} \left( {\lambda}^{2}+\gamma\,\delta \right) }}
&-
{\frac {2\,t{\lambda}^{2}x_{{t}}+2\,t{\lambda}^{2}\gamma\,{x}^{2}+2
\,t{\lambda}^{2}\delta-x\alpha\,\delta+x\beta\,\gamma}{4\lambda\,x
 \left( {\lambda}^{2}+\gamma\,\delta \right) }}\end {array} \right] 
,\eea
 which satisfy the zero-curvature condition 
 \bea\fl\label{ZCCP2} \Omega[x]=D_\lambda U^1-D_tU^2+[U^1, U^2]\nn
\fl \qquad =\frac12\left(x_{tt}-\frac{(x_t)^2}{x}+\frac{x_t}{t}-\frac1t(\alpha x^2+\beta)-\gamma x^3-\frac{\delta}{x}\right)\left[ \begin {array}{cc} \,{\frac {-\lambda\,t}{x \left( {\lambda
}^{2}+\gamma\,\delta \right) }}&\,{\frac {\gamma\,t}{{\lambda}^{2}+
\gamma\,\delta}}\\\noalign{\medskip}\,{\frac {\delta\,t}{{x}^{2}
 \left( {\lambda}^{2}+\gamma\,\delta \right) }}&\,{\frac {\lambda\,
t}{x \left( {\lambda}^{2}+\gamma\,\delta \right) }}\end {array}
 \right]=0 .\eea
 
 It has been shown, \cite{Gromak1999}, that one can find solutions of Painlev\'e P3 under the restrictions  $\beta=\delta=0$ or $\alpha=\gamma=0$.  Consider the former case, $\beta=\delta=0.$ With these restrictions  (\ref{P3}) admits a point symmetry given in evolutionary form by 
 \be \vec{w}_{\Q_1}=(x+tx_t)\frac{\partial}{\partial x}.\ee
 In the case $\alpha=\gamma=0$, the point symmetry of the reduced equation is
 \be \vec{w}_{\Q_2}=(x-tx_t)\frac{\partial}{\partial x}.\ee
 Below, we consider the two surfaces defined by these symmetries, which we index by a superscript indicating the considered integrable case, 
\bea  A=pr\vec{w}_{\Q_1}(U^1)=\frac{{\gamma}^{2}t{x}^{2}+ \left( \gamma+\alpha \right) x+\gamma\,tx_{{t}}}2\left[ \begin {array}{cc}1 &0\\\noalign{\medskip}0&-1\end {array} \right]\in sl(2,\mathbb{R}), \nn
B= pr\vec{w}_{\Q_1}(U^2)=\frac{{\gamma}^{2}t{x}^{2}+ \left( \gamma+\alpha \right) x+\gamma\,tx_{{t}}}2 \left[ \begin {array}{cc} t\lambda^{-1}&-{\frac {x+tx_{{t}}+
\gamma\,t{x}^{2}}{x{\lambda}^{2}}}\\\noalign{\medskip}0&-t\lambda^{-1}\end {array} \right] \in sl(2,\mathbb{R}),
\nn
I(F^{\Q_1})=\frac{\left({\gamma}^{2}t{x}^{2}+ \left( \gamma+\alpha \right) x+\gamma\,tx_{{t}}\right)^2}{4\lambda^2}\left(\lambda dt+td\lambda\right)^2,\nn
 N(F^{\Q_1})=\Phi^{-1}\left[\ba{cc}0 &1\\0 &0\ea\right]\nonumber\Phi, \eea
and in the second case
\bea A=pr\vec{w}_{\Q_2}(U^1)=\frac{\delta\,tx_{{t}}-x\delta+\beta\,x+{\delta}^{2}t}{2x^2} \left[ \begin {array}{cc} 1&0\\\noalign{\medskip}0&-1\end {array} \right]\in sl(2,\mathbb{R}), \nn
B=pr\vec{w}_{\Q_2}(U^2)=\frac{\delta\,tx_{{t}}-x\delta+\beta\,x+{\delta}^{2}t}{2x^2}\left[ \begin {array}{cc}t\lambda^{-1}&0
\\\noalign{\medskip}-{\frac {tx_{{t}}-x+\delta\,t}{{\lambda}^{2}{x
}}}&-t\lambda^{-1}\end {array} \right]\in sl(2,\mathbb{R}), \nn I(F^{\Q_2})=\frac{(-\delta\,tx_{{t}}+x\delta+\beta\,x-{\delta}^{2}t)^2}{4x^2\lambda^2}\left(\lambda dt+td\lambda\right)^2,\nn
N(F^{\Q_2})=\Phi^{-1}\left[\ba{cc}0 &0\\1 &0\ea\right]\Phi\nonumber.\eea
As in several of the previous cases, in the moving frame defined by conjugation by the (non-constant) wave function $\Phi$, the surfaces are contained in planes with isotropic normal vectors. Note that the Killing form on the tangents to the surfaces is degenerate, i.e. the determinants of the first fundamental forms are zero, however the tangent vectors are linearly independent and so the immersion functions define surfaces instead of curves. 

\section{Concluding remarks}
The main objective of this paper is to extend the applicability of the Fokas-Gel'fand procedure for constructing explicit soliton surfaces associated with integrable ODEs admitting zero-curvature representation. The most important advantage of this method is that it gives effective tools for constructing certain classes of soliton surfaces in a systematic way. It is much simpler and faster than traditional methods and its effectiveness has been demonstrated by the results obtained in section 5 for Painlev\'e equations P1-P3. At this point, we can summarize our approach for  constructing the immersion function for 2D-surfaces immersed in Lie algebras. 

\begin{itemize}
\item[1.] We have provided  a general framework for soliton surfaces whose Gauss-Mainardi-Codazzi equations are integrable deformations of the ZCC. This characterization includes three sub-cases, namely integrable PDEs in ZCC form, integrable ODEs in Lax representation and in ZCC form (see sections 2 and 3). 
\item[2.] We have provided a complete classification of all admissible generalized symmetries of the ZCC using the associated LSP. As a result, we have shown that any 2D-surface immersed in a Lie algebra is associated with a generalized symmetry of the ZCC (see section 2 Proposition 1 ''Extracting symmetries from a surface"). As a corollary, we have shown that all generalized symmetries of the ZCC can be written in terms of a gauge function on jet space. 
\item[3.] We have considered in depth the case of ODEs admitting Lax pairs in zero-curvature form and their associated surfaces, defined either explicitly by their tangent vectors or, when possible, from an explicit integrated form. 
\item[4.] To illustrate these theoretical considerations, we consider Lax pairs for Painlev\'e equations in zero-curvature form whose potential matrices take values in the Lie algebra $sl(2,\mathbb{R})$. For the considered equations,  we explore some geometric characteristics of the surfaces. These include the first and second fundamental forms, whose coefficients are not independent but subject to the differential constraints of the Painlev\'e equations P1, P2, and P3. The fundamental forms were then used to construct expressions for the  Gaussian and mean curvatures for the surfaces; in each case the Gaussian and mean curvatures are functionally independent. We also computed the umbilical points of several of the surfaces and in each case, the points lie on curves determined by polynomial equations in the spectral parameter. It is interesting to note that several of the obtained surfaces lie in planes in the (non-constant) moving frame defined by conjugation by the wave function $\Phi.$ 
 In the case of the P2 equation, we also consider specific rational and Airy function solutions. In the first case, for rational solutions, we were able to solve the determining equations for infinitesimal generators of symmetries and consequently the corresponding surfaces  are expressed in terms of Bessel and confluent Heun functions. 
\item[5.] We have described the integrable cases of the P3 equation, their associated point symmetries and their soliton surfaces. We have shown that the tangent vectors to the surfaces are planes conjugated by the (non-constant) wave function $\Phi.$
\end{itemize}

Finally, it is worth noting that the proposed approach for constructing 2D-surfaces associated with integrable models can be used "in reverse" to address certain physical problems. Namely, it is sometimes the case that a 2D-surface is known in terms of functions appearing in the physical system for which analytical models are not yet fully developed. Using our approach (Proposition 1), it is possible to select an appropriate surface immersed in a Lie algebra and evaluate some geometric properties such as the induced metric, mean curvature, and the corresponding Willmore functional. A variational problem of this functional allows for the straightforward computation of the class of equations describing the physical problem. These applications and further theoretical issues will be explored in our future works. 

\ack  The authors thank Professor R. Conte  for helpful discussions on this topic. The work was supported by a research grant from NSERC of Canada. S Post acknowledges a postdoctoral fellowship provided by the Laboratory of Mathematical Physics of the Centre de Recherches Math\'ematiques, Universit\'e de Montr\'eal.

\end{document}